\documentclass[aps,prb,amsmath,amssymb,twocolumn,showpacs,superscriptaddress]{revtex4}
\usepackage{graphicx}
\usepackage{bm}
\usepackage{epsfig}
\usepackage{natbib}
\usepackage{color}

\begin{document}

\title{Counterintuitive Consequence of Heating in
Strongly-Driven Intrinsic-Junctions of
Bi$_{2}$Sr$_{2}$CaCu$_{2}$O$_{8+\delta}$
Mesas}

\author{C. Kurter}
\email[Corresponding author: ] {ckurter@umd.edu}
\affiliation{Materials Science Division, Argonne National
Laboratory, Argonne, IL 60439, USA} \affiliation{Physics Division, BCPS Department,
Illinois Institute of Technology, Chicago, IL 60616, USA}
\author{L. Ozyuzer}
\affiliation{Materials Science Division, Argonne National
Laboratory, Argonne, IL 60439, USA} \affiliation{Department of Physics, Izmir Institute of Technology,
TR-35430 Izmir, Turkey}
\author{T. Proslier}
\affiliation{Materials Science Division, Argonne National
Laboratory, Argonne, IL 60439, USA} \affiliation{Physics Division, BCPS Department,
Illinois Institute of Technology, Chicago, IL 60616, USA}
\author{J. F. Zasadzinski}
\affiliation{Physics Division, BCPS Department,
Illinois Institute of Technology, Chicago, IL 60616, USA}
\author{D. G. Hinks}
\affiliation{Materials Science Division, Argonne National
Laboratory, Argonne, IL 60439, USA}
\author{K. E. Gray}
\affiliation{Materials Science Division, Argonne National
Laboratory, Argonne, IL 60439, USA}

\date{\today}

\begin{abstract}

Anomalously high and sharp peaks in the conductance of intrinsic Josephson junctions in Bi$_{2}$Sr$_{2}$CaCu$_{2}$O$_{8+\delta}$ (Bi2212) mesas have been universally interpreted as superconducting energy gaps, but here we show they are a result of heating. This interpretation follows from a direct comparison to the equilibrium gap, $\mathit \Delta$, measured in break junctions on similar Bi2212 crystals. As the dissipated power increases with a greater number of junctions in the mesa, the conductance peak abruptly sharpens and its voltage decreases to well below 2$\mathit \Delta$. This sharpening, found in our experimental data, defies conventional intuition of heating effects on tunneling spectra, but it can be understood as an instability into a nonequilibrium two-phase coexistent state. The measured peak positions occur accurately within the voltage range that an S-shaped backbending is found in the {\it calculated} current-voltage curves for spatially {\it uniform} self-heating and that S-shape implies the potential for the uniform state to be unstable.

\end{abstract}


\maketitle

\section{Introduction}

Intrinsic Josephson junctions (IJJs) in the crystal structure of the high-temperature superconductor (HTS) Bi$_{2}$Sr$_{2}$CaCu$_{2}$O$_{8+\delta}$ (Bi2212) exist along the c-axis between each pair of neighboring CuO$_{2}$ bilayers, with the Bi$_{2}$Sr$_{2}$O$_{4}$ layers acting as the insulating tunnel barrier. Kleiner et al.~\cite{Kleiner92} described the novel features of such Josephson junction stacks and continuing excitement about research prospects has led to an exhaustive literature on the current-voltage characteristics, $I(V)$, of IJJs, especially on sculpted mesas.~\cite{Yurgens00, Yurgens99, Fenton02, Yurgens03, Bae05, Zhu, Suzuki99, Krasnov05, Krasnov00, Katterwe, Bae08, Anagawa, Suzuki00, Fenton03, Kurter09} While it has long been recognized that self-heating of such mesas is a potential problem, the observation of high, sharp conductance peaks has been interpreted as evidence for minimal thermal broadening, and the peaks were thus ascribed to the quasiparticle gap feature expected from a stack of superconductor-insulator-superconductor (SIS) tunnel junctions. Important conclusions about the magnitude and temperature-dependence of the superconducting gap, $\mathit \Delta(T)$, along with inferences about an extrinsic pseudogap in Bi2212 have been drawn based on IJJs exhibiting these sharp peaks.~\cite{Suzuki99, Krasnov05, Krasnov00, Katterwe, Bae08} It has also been argued that the sharp peaks are evidence that the c-axis tunneling process is strictly coherent (see Yamada et al.~\cite{Suzuki99}).

Here we demonstrate unambiguously that similar sharp conductance peaks are {\it not} a measure of the superconducting gap, but they are rather a consequence of strong self-heating. This counterintuitive conclusion is based on: independent measurements of the tunneling spectra (and $\mathit \Delta$) on the same (or similar) crystals; a systematic control of heating via mesa height (number of junctions, $N$, at constant mesa area); and realistic heating models.  We find that only the shortest mesa data look similar to the equilibrium tunneling $I(V)$, albeit with a voltage-dependent mesa temperature, $T_m$, with $T_m$$<$ $T_c$ at the gap voltage. Intermediate size mesas reach $T_c$ before the gap voltage, due to their greater dissipation, and the transition to the normal state causes a sharp rise in their $I(V)$ {\it without} the conventionally-held sign of heating, i.e., backbending in $I(V)$.  We explain the lack of backbending by instability into a nonequilibrium two-phase coexistence for the transition to the normal state.  A realistic {\it uniform-temperature} self-heating model shows backbending of the $I(V)$ is not eliminated by including the anomalously large quasiparticle scattering rate, $\mathit \Gamma(T)$, that is taken from scanning tunneling spectroscopy (STS) data for Bi2212.~\cite{Pasupathy} Although uniform self-heating cannot therefore explain our data, the experimental peaks occur in the voltage range of backbending in the calculated $I(V)$ where instability might be expected, e.g., into two-phase coexistence (normal and superconducting) with a non-uniform temperature profile across the mesa area.

\begin{figure}
\centering
\includegraphics [width=3.4in]{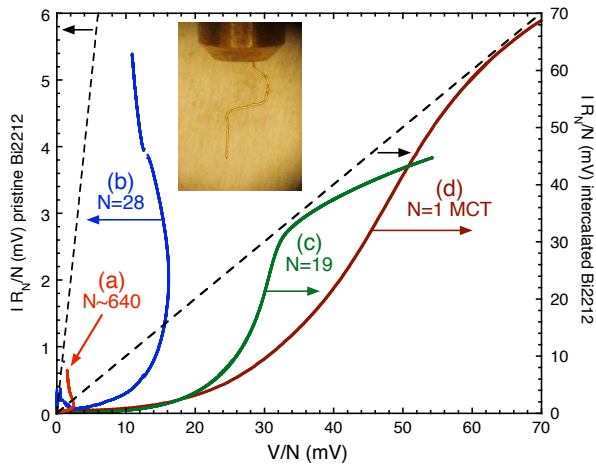}
\caption{(Color online) Comparison of tunneling data for 3 mesas and a single MCT SIS junction (d) at 4.2 K, except (a) which is at 26 K.  Note (a) and (b) use pristine Bi2212 and are plotted on the left hand axis while (c) and (d) use intercalated Bi2212 and are plotted on the right hand axis.  The dashed lines estimate the normal state resistance, $R_N$, in each case: (a) 12.8 $\Omega$; (b) 30 $\Omega$; (c) 460 $\Omega$; and (c) 15 $k\Omega$, The departure from this line at higher voltages, especially noticeable in (c), is caused by the energy dependence of the normal-state density-of-states and is also seen in the MCT SIN data displayed in Fig.~4.  Curve (a) is a 300x60 $\mu m^2$ THz emitting mesa of thickness 1 $\mu m$ ($N$$\sim$ 640). Inset: the 100-$\mu m$ soft Au wire used to contact the Au film atop the 10x10 $\mu m^2$ mesas in (b), (c) and in Fig.~2. The hook shape is to minimize contact force (damage) from the sharpened tip. }
\label{Fig1}
\end{figure}

Earlier studies~\cite{Yurgens03, Bae05, Zhu, Suzuki99, Krasnov05, Krasnov00} made reasonable conclusions about HTS properties based on data {\it only} from Bi2212 mesas. Such studies often reported $I(V)$ similar to Fig.~1(c) which displays a superconducting {\it gap-like} feature. However, a different story emerges when one compares these data to mechanical-contact tunneling (MCT),~\cite{Zasadzinski02, Ozyuzer00, Zasadzinski01, Miyakawa, Ozyuzer02, Ozyuzer05, Zasadzinski06} e.g., the SIS of Fig.~1(d), or STS.~\cite{Pasupathy, Renner, DeWilde, Pan, Dipasupil} Heating is a smaller problem for MCT and STS as junctions usually exhibit much lower dissipation due to: higher specific resistance; smaller areas; and improved heat removal as they consist of only one dissipating junction, rather than a mesa stack. If one compares published data on Bi2212 mesas to these traditional single junction methods, the only mesas that show similar behavior, like the spectral dip feature,~\cite{Zasadzinski02, Ozyuzer00, Zasadzinski01, Miyakawa, Ozyuzer02, Zasadzinski06, Renner, DeWilde} have greatly minimized heating effects. These latter mesas use short stacks ($N$$\sim$ 10) with either: intercalated Bi2212~\cite{Yurgens99, Choy} to reduce the dissipative c-axis quasiparticle conductance; ultra-small areas~\cite{Zhu} ($<$1 $\mu m^{2}$) to improve heat removal; or 60 ns current pulses~\cite{Fenton02, Anagawa, Suzuki00, Fenton03} to minimize the buildup of mesa temperature, $T_m$.

\begin{figure}
\centering
\includegraphics[width=3.4in]{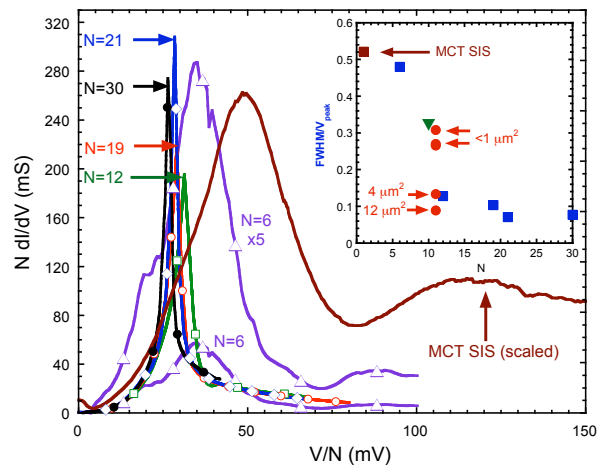}
\caption{(Color online) Conductance per junction, $NdI/dV$, for each intercalated mesa compared to scaled data for an intercalated MCT junction ($N$= 1). Inset: ratio of FWHM to the peak voltage, $\beta$, for each, together with those calculated for the data of Yurgens et al.~\cite{Yurgens99} - inverted triangle - and Zhu et al.~\cite{Zhu}- filled circles.}
\label{Fig2}
\end{figure}

Interest in Bi2212 IJJ stacks has been further peaked by the recent demonstration of emission of significant THz-wave radiation.~\cite{Ozyuzer07} In order to achieve the needed THz cavity resonance and high emission power, large mesa volumes were used and those are at odds with conventional trend in Bi2212 mesa research to reduce heating, being up to $10^{4}$-$10^{6}$ times larger in volume.  Thus the ongoing quest to understand non-equilibrium (NE) and self-heating effects in Bi2212 mesas is clearly front-and-center.

A stack of IJJs, in a mesa sculpted on a Bi2212 single crystal, offers a unique possibility for spectroscopy and exotic device arrays using well-defined, uniform tunnel barriers that are stable over the entire temperature range. However, the poor thermal conductivity, very large current density and close proximity of multiple neighboring junctions in the crystal structure of Bi2212 have led to significant concerns about heating. A striking example of self-heating is found in large volume mesas as a backbending of the $I(V)$, e.g., Fig.~1(a) and Fig.~1(b), occurring at voltages far below the gap voltage, 2$\mathit \Delta/e$. This backbending was shown~\cite{Kurter09,Ozyuzer07} to be self-heating that is dominated by the particular temperature dependence of the sub-gap quasiparticle c-axis resistivity, $\rho_c(T_m)$, rather than $\mathit \Delta(T_m)$ as was found previously in low-$T_c$ superconductors and explained by NE quasiparticle injection and/or self-heating effects.~\cite{Gray, Yeh} Compared to Fig.~1(a and b), the voltages of the sharp conductance peaks in our intercalated mesas, Fig. 1(c), are much closer to 2$\mathit \Delta$, so the effect of self-heating on $\mathit \Delta$ will not be negligible.

A basic question is whether an $I(V)$ curve such as Fig.~1(c), which exhibits a {\it gap-like} feature, is in fact a measurement of $\mathit \Delta(T)$. One of the main results of the present work is the establishment of straightforward analytical methods that allow such a question to be answered.  First we establish heating effects are present in Bi2212 mesas by directly comparing their $I(V)$ to those of single MCT junctions, using similarly made and doped crystals to eliminate uncertainties in sample quality or doping level. In Fig.~2, we compare the differential conductance per junction, $NdI/dV$, for intermediate-size mesas (10x10 $\mu m^{2}$ by 12-60 nm height, i.e., $N$= 6-30) with the MCT junction of Fig.~1(d), all taken on nominally identical intercalated crystals.  The differences are qualitatively and quantitatively unmistakable. Based on these data, we propose a figure-of-merit for IJJ stacks that is a quantitative measure of conductance peak broadening, $\beta$, given by the ratio of the peak's full width at half-maximum (FWHM) to its position, $V_{peak}$, such that $\beta$=FWHM/$V_{peak}$.  A summary of $\beta$-values for these data and those of others~\cite{Yurgens99, Zhu} are shown in the inset of Fig.~2. It is clear that when $\beta$$\leq$ 0.15 the high bias spectral dip feature (a definitive signature of superconductivity in Bi2212) is absent in our data and that of Zhu et al.~\cite{Zhu}  We will show below that the observed, sharp {\it gap-like} feature when $\beta$$\leq$ 0.15 is not a measure of $\mathit \Delta$, but rather marks the transition of the mesa into the normal state via two-phase coexistence that is due to self-heating.  Applying this criterion to previously published work, it is evident that some mesa studies, which presumed to be measuring $\mathit \Delta(T)$, were likely observing self-heating effects instead.

It is counterintuitive that the data for intermediate size mesas ($N$= 12 to 30) become sharper with greater heating (i.e., larger $N$) as this does not reflect the usual smearing-out of the junction $I(V)$ due to heating. This is especially true here since smearing should be severe in view of the anomalously large quasiparticle scattering rate, $\mathit \Gamma(T)$, for Bi2212 near $T_c$.~\cite{Pasupathy} Thus our conclusion that conductance peaks result from heating defies intuition and its justification is the primary purpose of this paper. Our contention is based on a self-heating induced transition into the normal state through two-phase coexistence. We will also show that the {\it gap-like} feature shifts to lower voltages {\it with increasing bath temperature}, and this effect mimics reports in the mesa literature that are interpreted as closing of  $\mathit \Delta(T)$ at $T_c$. However, despite this feature not being $\mathit \Delta$, its disappearance at $T_c$ is guaranteed in our self-heating scenario because there can then be no transition {\it into} the normal state if $T$ is above $T_c$.

\section{Experiment}

Single crystals of Ca-rich Bi$_{2.1}$Sr$_{1.4}$Ca$_{1.5}$Cu$_{2}$O$_{8+\delta}$, were grown by a floating zone technique. Intercalation of HgBr$_{2}$ occurred upon heating these crystals in air with excess HgBr$_{2}$ gas at 230 $^\circ$C for 16 hours and x-ray diffraction confirmed the c-axis lattice constant increased from 1.531 nm to 2.151 nm, as found previously.~\cite{Choy} The intercalated crystals exhibited T$_c$$\sim$ 74 K from magnetization and $\mathit \Delta$$\sim$ 24 meV from MCT, indicating they are likely overdoped.~\cite{Ozyuzer00} Intercalation of HgBr$_{2}$ between the BiO layers reduces the specific dissipation at fixed voltage by thickening the Bi$_{2}$Sr$_{2}$O$_{4}$ tunnel barrier to obtain an order-of-magnitude decrease in the c-axis conductance. We can fine-tune self-heating by of a single control parameter, the stack height, since total heating power is proportional to $N$ for a constant mesa area.

Intercalated crystals were cleaved, sputter coated with gold and Ar-ion beam etched~\cite{Kurter05} into arrays of 10x10 $\mu m^{2}$ mesas using photolithography. Our MCT apparatus, described by Ozyuzer et al.~\cite{Ozyuzer98}, is also used to contact the gold film atop the mesa with a soft, 100 $\mu$m- diameter gold wire that is bent in a hook-shape to minimize the contact force and any potential damage to the mesa (see inset of Fig.~1). This wire is sharpened to a diameter of 5-10 $\mu m$ at the end touching the mesa and invariably the tip contacted a single mesa of the array. The multiple-sweep $I(V)$ in Fig. 3 shows the Josephson current and the number of quasiparticle branches~\cite{Kleiner94} corresponding to the number, $N$, of IJJs in the mesa.

\begin{figure}
\centering
\includegraphics[width=3.4in]{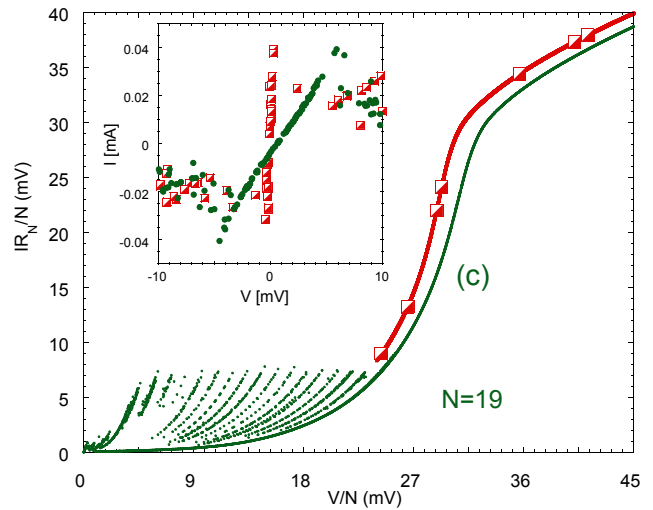}
\caption{(Color online) Details of curve (c) of Fig.~1 showing all quasiparticle branches and the $I(V)$ obtained after subtracting the initial SIN junction (half-open symbols). Inset: shows this subtraction accounts for the slope of the zero-voltage dc Josephson branch. }
\label{Fig3}
\end{figure}

Single-junction methods were employed for comparison with the IJJ spectra. Both superconductor-insulator-normal metal junctions (SIN) and SIS break junctions were obtained on an intercalated crystal by MCT using a much thicker gold wire with a blunt tip. After collecting SIN data, shown in Fig.~4, a hard contact is used to micro-cleave the underlying crystal leaving a chip of intercalated Bi2212 on the Au tip for subsequent SIS junctions,~\cite{Ozyuzer00} as shown in Fig.~1(d). Since heating is virtually eliminated for MCT, these data provide the equilibrium properties for interpreting the Bi2212 mesa data. The mesa's top-contact gold film forms an SIN junction in series with the IJJ stack: it is seen as a finite resistance (slope) for the zero-bias Josephson supercurrent (solid symbols of the inset to Fig.~3). This mesa resistance {\it decreases} with temperature in a manner that is only consistent with an SIN junction. To correct for this, our $I(V)$ from MCT in the SIN configuration is used to subtract the top-contact SIN voltage from the measured total voltage, for every current. The result (open symbols in both Fig.~3 and inset) is the $I(V)$ of the IJJ SIS stack alone and a numerical derivative generates the $dI/dV$ of Fig.~2. The biggest SIN correction yields a $10\%$ lower peak voltage, but our qualitative conclusions would be unaltered by neglecting this correction.

The telltale signs of heating in the $dI/dV$ of Fig.~2, even in the absence of backbending in $I(V)$, are: (1) anomalously high and narrow peaks in the taller mesas; (2) significantly reduced peak voltages from the MCT value; and (3) weak ($N$=6) or absent dip-hump-feature (DHF) that is seen above the peak in the MCT data (at $\sim$ 80 and $\sim$ 120 mV in the MCT data of Fig. 2). The DHF is a universal feature of all previous MCT,~\cite{Zasadzinski02, Ozyuzer00, Zasadzinski01, Miyakawa, Ozyuzer02, Zasadzinski06} STS,~\cite{Pasupathy, Renner, DeWilde, Pan, Dipasupil} and angle-resolved photoemission~\cite{Campuzano} studies as well as IJJ data~\cite{Yurgens99, Zhu, Anagawa} with reduced heating.

To validate the generality of our results, we note that intercalation of HgBr$_{2}$ has little effect on the DOS of the Cu-O bilayers as revealed by the similarity of the SIN MCT data in Fig.~4, with and without intercalation. Both curves are fit by a d-wave DOS:~\cite{Won}

\begin{equation}
{N}_{s }(E,k)= Re\left(\frac{E- \iota  \mathit \Gamma}{\sqrt{ ({E-\iota  \mathit \Gamma})^2-{\mathit \Delta(k)}^2}}\right )
\end{equation}
where $E$ is the quasiparticle energy, $\mathit \Delta(k)$= $\mathit \Delta _{0}cos(2 \phi)$,  $\phi$ is polar angle in k-space and $\mathit \Gamma$ is assumed to arise from the quasiparticle scattering rate.  The sub-gap conductance is well modeled by Eq.~1 and the fit, shown for the intercalated crystal, gives $\mathit \Delta _{0}$= 24 meV and  $\mathit \Gamma$= 0.55 meV while the oxygen overdoped pristine Bi2212 crystal fit (not shown) yields $\mathit \Delta _{0}$= 26 meV and  $\mathit \Gamma$= 0.6 meV, and both use the same momentum-averaging parameter.~\cite{Ozyuzer00} At higher voltages, Eq.~1 is inadequate as it misses the actual normal-state conductance and the $E$-dependence of  $\mathit \Delta$: the latter produces the DHF~\cite{Eschrig} that is seen more prominently at negative voltages in Fig.~4.

\begin{figure}
\centering
\includegraphics[width=3.2in]{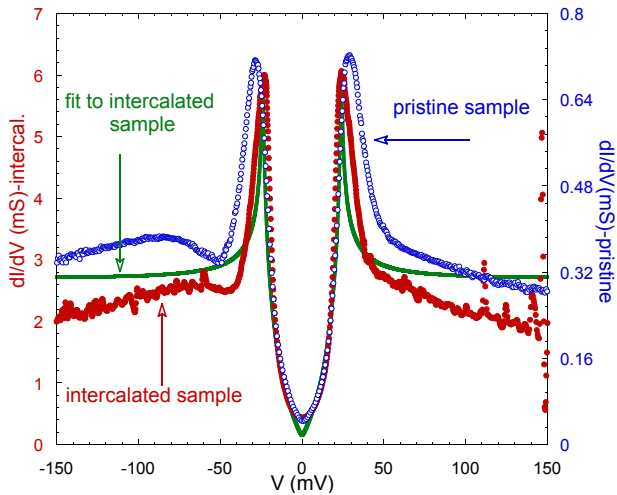}
\caption{(Color online) The $dI/dV$ from MCT for intercalated (solid circles) and pristine (open circles) Bi2212 are very similar and for subgap voltages they fit the momentum-averaged~\cite{Ozyuzer00} d-wave BCS model~\cite{Zavaritsky} (solid line). Above $\mathit \Delta$, the energy dependence of the normal-state density-of-states is clearly seen.}
\label{Fig4}
\end{figure}

The tunneling conductance of an SIS MCT break junction in Fig.~2 exhibits both Josephson (small peak at zero bias) and quasiparticle tunneling.  Because SIS junctions are less sensitive to thermal broadening, their conductance {\it peak} voltages are an excellent measure of 2$\mathit \Delta$. Thus $\mathit \Delta$= 24 meV from the SIS peaks matches that of the SIN junction fit of Fig.~4. The dip feature located at 80 mV is more readily visible in the SIS data but can be traced consistently~\cite{Zasadzinski02} to the feature at -50 mV in the SIN DOS of Fig.~4. For example, using the difference between the dip minimum and the peak voltage as a crude estimate, the strong coupling boson energy~\cite{Zasadzinski06} gives approximately 25 meV and 30 meV for the SIN and SIS junctions respectively, a reasonable consistency. Notably, data well above the peaks coincide for all mesas with $N$= 12-30 in Fig.~2, and the lack of DHF implies that all vestiges of superconductivity are gone. Then $NdI/dV$ at the dip/hump for $N$= 6 is clearly seen to fall below/above that normal state value as expected for strong-coupling self-energy effects linked to superconductivity, and in agreement with the MCT data.~\cite{Zasadzinski02, Ozyuzer00, Zasadzinski01, Miyakawa, Ozyuzer02, Zasadzinski06}

While the conductance peaks in Fig.~2 for mesas with $N$$>$ 6 are reminiscent of the superconducting coherence peak at 2$\mathit \Delta$, that interpretation is incorrect. First, the peak is at a voltage that is significantly smaller than 2$\mathit \Delta/e$, the equilibrium value from MCT. Furthermore, as the mesa height, and thus dissipation, increases, the peak voltage per junction decreases and that trend is suggestive of heating. Note that the ratio, $\beta$, of a peak's full-width-half-maximum to its voltage (inset of Fig.~2), is a small, almost constant value for $N$= 12 to 30, while it is considerably larger for the $N$= 6 mesa and the MCT data that both exhibit a well-defined DHF. The data reveal an abrupt change in $\beta$  by nearly a factor of 6 occurring between $N$= 6 and 12.  This large change in $\beta$ signals an important crossover from near-to-equilibrium superconducting properties to severe self-heating. To make this result more general, the inset of Fig.~2 includes data from Yurgens et al.~\cite{Yurgens99} and Zhu et al.~\cite{Zhu} which are on mesas with $N$= 10-11. The Zhu data are particularly useful as they utilize a different independent control of self-heating via the mesa area.  Nevertheless, a similar abrupt change of $\beta$ versus mesa area is observed in the inset of Fig.~2, and it implies that smaller area mesas better utilize the lateral spread of heat in the underlying crystal (suggested in Fenton et al.~\cite{Fenton03}). For $\beta$$>$ 0.3, both the Yurgens and Zhu data observe the characteristic DHF but for $\beta$$<$ 0.15 this feature disappears.

Note that recent STS data~\cite{Dipasupil} show that the DHF disappears as T approaches $T_c$, so its absence likely indicates that $T_m$$\geq$ $T_c$ after $NdI/dV$ returns to the normal-state value for $N$= 12-30 in Fig.~2.  In addition, the onset of normal-state behavior corresponds to the same dissipated power, of $\sim$ 1.05 mW, for $N$= 12-30.  For $N$= 6, the DHF occurs at power levels between $\sim$ 0.5 and 1 mW, while for $N$= 12 a very faint feature is seen at $\sim$ 0.9 mW.

To summarize this section, we have shown that sharp conductance peaks ($\beta$$<$ 0.15) in taller mesas are not a measure of $\mathit \Delta(T)$ despite their {\it gap-like} appearance. Rather, their origins are the sharp rise in $I(V)$ that represents {\it transitions of the mesa into the normal state}, and these are found over a range of $N$. The following section presents a detailed analysis of heating as the cause of this transition.

\section{Heating Analysis}

Ideal BCS superconductors (s-wave) should always exhibit backbending in their SIS junction $I(V)$ because $\mathit \Delta$ decreases as the effective T inevitably increases with current (power).  However, for weak heating, backbending may be absent due to broadening of the $I(V)$ by the intrinsic quasiparticle scattering rate, $\mathit \Gamma$, or by a spatially inhomogeneous $\mathit \Delta$. But if heating is significant for V$ $$\sim$ 2$\mathit \Delta(T_B)/e$, where $T_B$ is the bath temperature, backbending is observed.~\cite{Yeh} For Bi2212, several factors could reduce the likelihood of observing backbending.  The d-wave DOS (Eq.~1) does not produce as sharp a jump in $I(V)$ when $\mathit \Gamma$ is zero, but also the large $\mathit \Gamma$, that increases dramatically as $T_c$ is approached, might eliminate backbending even for strong heating and thus reproduce our mesa data. We test this latter possibility in {\it Section III.A} by calculating the $I(V)$ for Bi2212 mesas using experimental $\mathit \Gamma$ values in Eq.~1, but we always find backbending for $N$$>$ 12. Note however, that within the voltage range for backbending, the S-shaped $I(V)$ allows three different currents, so the situation is potentially unstable to a spatial breakup into hot spots $(T_{hs}$$\geq$ $T_c)$ and superconducting regions.  The resulting two-phase coexistence, described in {\it Section III.B} below, can qualitatively explain the absence of backbending in our experimental $I(V)$.  In addition, our sharp conductance peaks, that would signal the transition into the normal state through the growth of hot spots, occur for all $N$ accurately within the unstable, backbending voltage range of our $I(V)$ calculated in {\it Section III.A}.

\subsection{Uniform Heating Calculation}

The electrical power dissipated within the tunnel junctions of a mesa results in increases of the average excitation energies of the electron and phonon systems.  The short relaxation times found at temperatures of 50-100 K allow one to define an effective mesa temperature, $T_m$, for these steady-state electron and phonon distribution functions. It has been shown~\cite{Kurter09} that Newton's law of cooling,~\cite{Zavaritsky}
\begin{equation}
 { T}_{m }= { T}_{B }+ \alpha P
\end{equation}
represents a good approximation for Bi2212 mesas.  Here $P=IV$, $V$ is the voltage across the stack of SIS junctions, $\alpha$ is the effective thermal resistance and $T_B$ is the bath temperature.  For spatially uniform heating one would always probe a near-to-equilibrium system that would be, however, at a current (or power) dependent effective temperature given by Eq.~2. Thus our $I(V)$ do not represent any constant-temperature equilibrium $I_{eq}(V,T)$ found with negligible dissipation (e.g., by MCT or STS probes).  To emulate our mesa $I(V)$ data, we need to self-consistently determine for each $I$ and $V$ an effective $T_m$, for which $I$, $V$ and $T_m$ satisfy both Eq.~2 and the equilibrium $I_{eq}(V,T_m)$.

We determine the mesa $I(V)$ for a stack of $N$ junctions by a straightforward solution of these two independent relations among the values of $I$, $V$ and $T_m$. We generate equilibrium $I_{eq}(V,T)$ from standard tunneling theory of SIS junctions using the DOS of Eq.~1 and experimental values of  $\mathit \Delta(T)$ and  $\mathit \Gamma(T)$ from STS data.~\cite{Pasupathy} These STS data are similar to the MCT example of Fig.~4, but over a wide range of $T$.  That study~\cite{Pasupathy} used one unintercalated Bi2212 crystal ($T_c$$\sim$ 74 K) and found numerous data sets that correspond to the observed variation of the local properties across the crystal surface.  We used three of these data sets, labeled according to their low temperature gaps, $\mathit \Delta_1$, $\mathit \Delta_2$ and $\mathit \Delta_3$= 22 meV, 23 meV and 25 meV, respectively.  These gaps and $T_c$ are close to those reported here, and are thus representative of our intercalated crystals (recall that Fig.~4 shows the similarity of pristine and intercalated Bi2212 MCT data). The STS data sets, $\mathit \Delta(T)$ and $\mathit \Gamma(T)$, determine the quasiparticle DOS from Eq.~1, which is then convoluted with itself to generate the equilibrium  $I_{eq}(V,T)$, shown in Fig.~5a for the STS data set $\mathit \Delta_1$, $\mathit \Gamma_1$.

To illustrate this method, we consider a fixed current, $I$. Then the allowed mesa voltages are represented by two independent, single-valued functions for each fixed $I$: (1) $V_1(I,T_m)$= $NV_{eq}(I,T_m)$, where $V_{eq}(I,T_m)$ is the inversion of the calculated $I_{eq}(V,T_m)$ and it is generally a decreasing function of $T_m$; and (2) $V_2(I,T_m)$= $(T_m-T_B)/ \alpha I$, which follows from Eq.~2 and is an increasing function of $T_m$. The intersection of the two functions is the self-consistent solution for the mesa voltage, $V$, and temperature, $T_m$, for each value of $I$.  This procedure, repeated for all values of $I$, leads to the full $I(V)$. Although these solutions represent spatially uniform heating, the effective $T_m$ will be current (or power) dependent. Those $T_m$ are shown for a few curves in Fig.~5b that illustrates our model calculations for $N$= 12 to 30 using the data sets of Pasupathy et al.~\cite{Pasupathy} (labeled $\Gamma_1$ and  $\Gamma_3$, while $\Gamma_2$ data, not shown, fall between them).

That all data sets for $N$$>$ 12 show backbending implies that strong quasiparticle damping, $\mathit \Gamma(T)$, near $T_c$ is insufficient to smear it out.  To test the effect of $\mathit \Gamma$, a fourth curve is shown in Fig.~5b for $N$= 19 that uses $\mathit \Gamma(T)$= 2$\mathit \Gamma_3(T)$  and the degree of backbending is reduced, but not eliminated.  Although this backbending conflicts with our measured $I(V)$, our measured peaks of Fig.~2 occur at the same voltages as the backbending region or the sharp upturn in the calculated $I(V)$, for all $N$. This agreement is demonstrated in Fig.~6 for various $N$ and for data sets  $\mathit \Delta_1$, $\mathit \Delta_2$ and $\mathit \Delta_3$.  When the calculated curves show backbending, we plot both extrema to give a quantitative measure of the region of instability. The optimum agreement in Fig. 6 was found by setting  $\alpha$= 110 K/mW.

Reasonable constraints on $\alpha$ can be estimated from Fig.~2 and Pasupathy et al.~\cite{Pasupathy}  Recall that DHF disappears as $T$
approaches $T_c$,~\cite{Dipasupil} so its absence just above the sharp conductance peaks (for $N$= 12-30) likely indicates that these mesas are in the normal state at that voltage.  These points,  for $N$= 12-30, are all at a power of 1.0-1.1 mW.  Since the lower bound on the effective $T_m$ for this normal state, without a DHF, is $T_c$= 74 K, the lower bound on $\alpha$ is $\sim$ 70 K/mW  since $T_B$$\sim$ 4 K.  However, the lack of any residual structure in the conductance data above the sharp peak implies that we may better approximate the effective $T_m$ for this normal state as that for which the gap fluctuations (or pseudogap effects) above $T_c$ become negligible. From the STS data on pristine Bi2211~\cite{Pasupathy} used in our analysis, this would be $\sim$ 84 K.  But in addition, intercalated Bi2212 presents a much higher anisotropy so fluctuations likely persist to higher temperatures and  $\alpha$= 110 K/mW is not unrealistic.  In any case, the accurate $N$-dependence of Fig.~6 is a more rigorous test of the model than the precise value of $\alpha$ .

Unfortunately, the backbending observed in the model calculations of Fig.~5b is totally absent in the experimental curves. Thus the uniform heating model is incomplete, even though it quantitatively captures (Fig. 6) the decrease in peak voltage of the mesa data as a function of $N$. This discrepancy can be rectified if one recognizes that intrinsic backbending is a potentially unstable situation that can lead to two-phase coexistence.

\subsection{Two-Phase Coexistence Across the Mesa Area}

In order to understand the very narrow peaks with their absence of backbending, we postulate that the transition to the normal state can be via two-phase coexistence across the mesa area. One part of the mesa area could exhibit a smaller-than-average current density and power dissipation to remain superconducting (T$<$$ T_c$) while another portion, a hot spot, supports the normal state ($T_{hs}$$\geq$ $T_c$)  with a compensating larger-than-average current (power) density.  The instability into an inhomogeneous state follows from the intrinsic S-shaped $I(V)$, calculated above for uniform heating, that exhibits three different currents for any constant $V$ in the backbending region.  The Au film atop the mesa and the high transport anisotropy of Bi2212 assures a uniform $V$ across the entire mesa area, even with two-phase coexistence. The highest current solution represents the normal state, the lowest current solution is the superconducting state and the middle one is likely unstable.  Once in this non-uniform state, further increases in current require a larger hot spot, suggesting that it would then propagate across the entire mesa area as the current increases. The power density needed to maintain the hot spot at $T_c$ should be weakly dependent on its size, so the increase in current would occur at roughly constant voltage, in agreement with our sharp conductance peaks (Fig.~2).

Such a NE driven state of inhomogeneous current with a sharp conductance peak has been previously documented in low-$T_c$ superconducting junctions,~\cite{Gray} and the occurrence of hot spots $(T_{hs}$$\geq$ $T_c)$ with a size proportional to the current has been reported recently in very large Bi2212 mesas.~\cite{Wang} However, unlike our data, these cases exhibit a jump in the $I(V)$ to lower voltage and higher current as an initial hot spot nucleates. Rationalizing this difference requires a digression into the thermal boundary conditions for various mesa configurations.  That an increase in current leads to a larger hot spot is an integral part of any two-phase model and is justified by the data of Gray et al. and Wang et al.~\cite{Gray, Wang} The relevant point is whether the mesa voltage increases or decreases as the hot spot expands.

\begin{figure}
\centering
\includegraphics[width=3.4in]{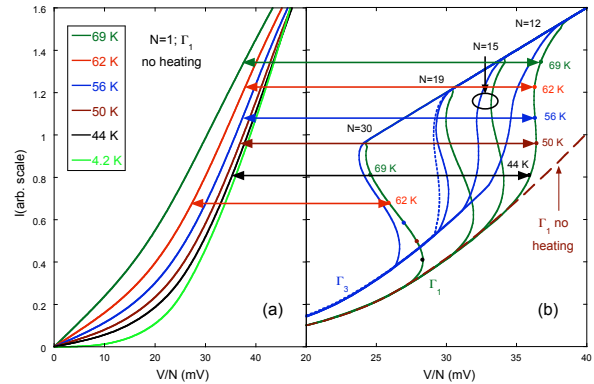}
\caption{(Color online) Numerical simulations of $I(V)$ using  $\mathit \Delta(T)$ and $\mathit \Gamma(T)$ from reported fits to the experimental STS data.~\cite{Pasupathy} (a) equilibrium $I(V)$ for various fixed temperatures $T$, using  $\mathit \Delta_1$, $\mathit \Gamma_1$  from Pasupathy et al.~\cite{Pasupathy} (b) non-equilibrium $I(V)$ including self-heating for various $N$ and two of the data sets $\mathit \Delta_1$, $\mathit \Gamma_1$ and $\mathit \Delta_3$, $\mathit \Gamma_3$ taken from Pasupathy et al.~\cite{Pasupathy} (calculations for $\mathit \Delta_2$, $\mathit \Gamma_2$ lie between these).  Dashed curve for $N$= 19 shows effect of doubling the quasiparticle lifetime smearing, i.e., $\mathit \Gamma$= 2$\mathit \Gamma_3$.  For two curves the local uniform $T_m$ values are shown and they correspond to points on the equilibrium $I(V)$. This correspondence was used in Kurter et al.~\cite{Kurter09}  and Ozyuzer et al.,~\cite{Ozyuzer07} but in those cases only the linear, low-voltage portion of the equilibrium $I(V)$ were needed, e.g., Fig.~1(a).  See text for further details.  }
\label{Fig5}
\end{figure}

The data of Zhu et al.,~\cite{Zhu} who varied only the mesa area, are insightful (see Fig.~2 inset).  Decreasing the area led to less heating, and eventually, for mesa areas $<1$$\mu m^2$, they found close-to-equilibrium properties. This is most readily explained by a more effective lateral transport of heat in the underlying crystal as the mesa area decreases, and this idea was expressed by Fenton et al.~\cite{Fenton03} Translating this idea into a two-phase model, larger hot spots, because of their poorer lateral heat transport, require {\it less} power density to maintain $(T_{hs}$$\geq$ $T_c)$.  The power density is $(V/t)^2/\rho_{cNc}$, where $t$ and  $\rho_{cNc}$ are the mesa height and normal-state c-axis resistivity at $T_c$, respectively.  Thus as the current and hot spot size increase, the required voltage {\it decreases} since $t$ and  $\rho_{cNc}$ are constant.  This admits the possibility of jumps to a larger hot spot, with its lower voltage and higher current, since then such a jump can be consistent with the electrical bias load line. This scenario is presumably the explanation of the jumps seen in the large mesa~\cite{Wang} and our Fig.~1(b), while the boundary condition for the jumps in thin films is likely different.~\cite{Gray}

For the intercalated mesas reported here, heat removal through the underlying crystal is more problematic. The large HgBr$_{2}$ intercalant molecules should behave as phonon rattlers~\cite{Nolas} to reduce the thermal conductivity (ratios of $\geq$30 were found in Nolas et al.~\cite{Nolas}) compared to pristine mesas.  As such, the relatively low thermal resistance of our 100 $\mu m$-diameter, $\sim$ 0.5 cm long Au-wire ($\sim$ 2 K/mW) contacting the 40-nm Au film atop our mesas (Fig.~1 inset) is a more significant heat sink. If dominant, this heat sink would critically alter the heat-flow path from a model that considers the underlying crystal as the only heat sink. Then the hot spot should nucleate farthest from the (random) position of the Au-wire contact. As the hot spot grows toward the Au wire, the added regions in closer proximity to the Au-wire heat sink need a {\it larger} power density to maintain $T_{hs}$$\geq$ $T_c$.  Thus as the hot spot increases in size, both the total current and voltage will {\it increase}.  Then the differential resistance is always positive throughout the two-phase region, and that is in accord with our mesa data in Fig.~1(c) for intercalated Bi2212. This behavior occurs in spite of the S-shaped $I(V)$ for that mesa if the heating was homogeneous. Another way to visualize this transition to the normal state is to recognize that the Au-wire heat sink produces a thermal gradient across the mesa whenever power is dissipated.  The most remote points will reach $T_c$ first and that phase boundary will move continuously toward the heat sink as the current (power) is increased.

\begin{figure}
\centering
\includegraphics[width=3.4in]{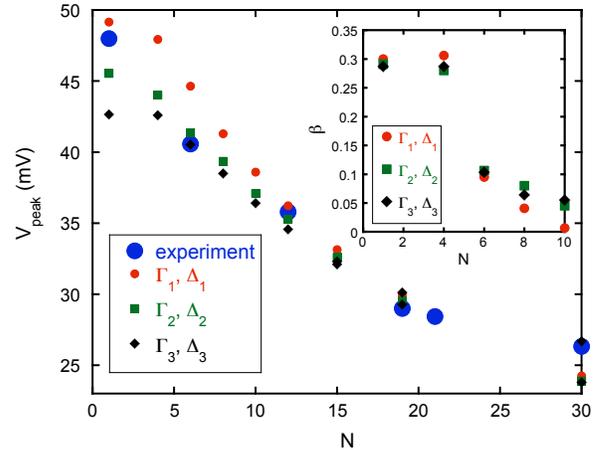}
\caption{(Color online) Excellent agreement is demonstrated between the experimental conductance peak voltage, $V_{peak}$, (large solid circles) and the {\it gap-like} features of the calculated $I(V)$ for each of the relevant STS data sets, ($\mathit \Delta_1$, $\mathit \Delta_2$ and $\mathit \Delta_3$) and a thermal resistance,  $\alpha$= 110 K/mW.  Examples of these calculations are shown in Fig. 5b.  Inset: ratio, $\beta$, of FWHM to $V_{peak}$ for the calculated curves that do not show backbending. The dramatic decrease between $N$= 4 and $N$= 6 mimics the experimental one seen in Fig.~2 (inset). }
\label{Fig6}
\end{figure}

The previous discussion explains how two-phase coexistence can eliminate backbending in intercalated mesas, but can we understand why Fig.~1(b) with the same mesa area and $N$= 28 behaves so differently?  In curve (b) there is a jump seen for $IR_N/N$$\sim$ 3.9 mV announcing the formation of a hot spot,~\cite{Wang} while the backbending part of the curve may be effectively explained by the temperature dependence of  $\rho_{cN}(T)$ in the regions outside the hot spot.~\cite{Kurter09}  The only significant difference is that mesa (b) is made on a {\it pristine, unintercalated} Bi2212 crystal.  This yields an order-of-magnitude larger power density for a given voltage per junction but more importantly, as mentioned above, a considerably higher thermal conductivity for the mesa and its underlying crystal than for the intercalated mesa (c). Therefore for the pristine Bi2212 mesa, curve (b), the thermal agenda may be set by the underlying crystal rather than the gold-point top contact (Fig.~1 inset).

For the $N$= 28 mesa of curve (b), the heat transfer coefficient is found to be  $\alpha$= 38 K/mW, by using the methods of Kurter et al.~\cite{Kurter09} and the full set of $I(V,T)$.  The above estimate of the conduction through the thin-Au film contact atop this mesa is much smaller ($\alpha$$\sim$ 110 K/mW).  Thus the underlying crystal is the predominant source of cooling for this unintercalated mesa.

The fraction size of mesa comprising the hot spot, $f$, can be reasonably estimated.  First, we find the contributions to the total current, $I_f$, from the hot spot's normal-state current, $fV_f/R_N$, and the quasiparticle current in the remaining part of the mesa, $(1-f)I_{qp}$, where $I_{qp}(V_f)$ is the lowest branch of curve (b) of Fig.~1. Thus $f$= $(I_f-I_{qp})/(V_f/R_N-I_{qp})$$\sim$ 0.26. Next, we estimate the power density needed to achieve the normal state, $T_N$, using the hot spot voltage, $V_f$, and the heat transfer coefficient, $\alpha \sqrt{f}$. Here $\sqrt{f}$ approximates the radius dependence of heat transfer through the underlying crystal~\cite{Fenton03} since the $\alpha$ determined above was for uniform heating of the {\it entire} mesa area. This analysis gives $f$$\sim$ $((T_N-T_B)R_N/\alpha V_f^2)^2$ and its evaluation for $f$= 0.26 implies a reasonable value of 84 K for the only adjustable parameter, $T_N$. This is consistent with superconducting fluctuations (or a pseudogap) found out to $\sim$ 85 K for unintercalated Bi2212.~\cite{Pasupathy}

\begin{figure}
\centering
\includegraphics[width=3.4in]{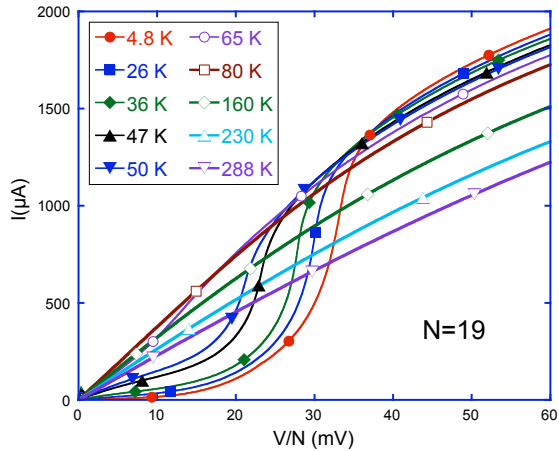}
\caption{(Color online) The bath temperature, $T_B$, dependence of the $I(V)$ for the N=19 intercalated mesa. The {\it gap-like} feature at $\sim$ 33 mV for $T_B$= 4.8 K is seen to decrease with increasing $T_B$ and completely disappear for $T_B$= 80 K.     }
\label{Fig7}
\end{figure}

Although providing a justification for the observed jump to be a hot spot, this calculation does not rule out its occurrence at a lower current in the backbending region of Fig.~1(b). From the above analysis a lower current would correspond to a smaller hot spot, i.e., a smaller $f$. The heating analysis by Fenton et al.~\cite{Fenton03} implies there is a minimum size to achieve $T$$\geq$ $T_N$ in the hot spot. To find it, we integrate their formula~\cite{Fenton03} for the steady-state temperature rise using the parameters of our hot spot and get a 5-$\mu m$ radius for $\delta T=80$ K, whereas the above analysis leading to $f$$\sim$ 0.26 implies a radius of 2.9 $\mu m$. To expect much better agreement may be unreasonable given our imprecise knowledge of the proper values to use for the mean thermal conductivity, extrapolation for $R_N$, $I_{qp}$, etc.

To summarize, the variety of behavior seen in Bi2212 mesas (Fig.~1) can be understood by a coherent heating model that includes two-phase coexistence for highly driven mesas and uniform heating for moderately driven mesas.

\section{Additional Features of the Experimental Data and Modeling }

Although incomplete, the uniform heating model accurately predicts (see Fig.~6) the measured conductance peak voltages of Fig.~2 as the voltage of the backbending instability for $N$$\geq$ 12.  This optimum agreement uses a reasonable value of the one free parameter, $\alpha$$\sim$ 110 K/mW, and thus provides convincing, supporting evidence that the sharp peaks ($\beta$$<$ 0.15) are a consequence of strong self-heating. Since the agreement in Fig.~6 covers the range from weak ($T_m$$\ll$ $T_c$) to strong self-heating, it is of interest to see if the calculation shows the abrupt change in  $\beta$ with $N$ that is found experimentally. The inset of Fig.~6 shows the $\beta$ values derived from calculations for small $N$ values for which backbending {\it has not yet developed}. Note that the abrupt decrease in the calculated $\beta$ between $N$= 4 and $N$= 8 is reasonably close to the experimental one (Fig.~2 inset).

This abrupt decrease in $\beta$  signals a crossover in how the tunneling conductance achieves its normal-state value: for large $N$ it is due to strong self-heating ($(T_m$$\geq$ $T_c)$)  while for small $N$, the mesa remains superconducting and $dI/dV$ merges with the normal-state conductance, $Y_N$, as it does in thermal equilibrium when $V$ equals the value of 2$\mathit \Delta(T_m)/e$, where $T_B$$<$ $T_m$$<$ $T_c$. This perspective comes from comparing the calculated $I(V)$ for $N$= 4 and $N$= 8. For $N$= 4, $T_m$ is only 38 K at the gap voltage, $V$= 2$\mathit \Delta(T_B)/e$= 48 mV, while for $N$= 8, $T_m$ reaches $T_c$= 74 K and the conductance reaches $Y_N$ at a lower V= 44 mV.  This analysis affirms our previous assertion.  Thus the model confirms what the data already told us and it distinguishes $\beta$ as an appropriate figure-of-merit for strong self-heating effects that mimic the superconducting gap feature in $I(V)$.

Our data, modeling and discussion, so far, have concentrated on $T_B$$\sim$ 4.2 K$\ll$ $T_c$. Figure~7 shows the dependence of $I(V)$ on $T_B$ for the mesa with $N$= 19, for which we have already established that its {\it gap-like} feature, shown for 4.2 K in Fig.~1(c), is due to two-phase coexistence caused by self-heating.  Upon increasing $T_B$, this feature shifts to lower voltages and disappears completely at T=80 K where a nearly linear $I(V)$ is found out to $V$$\sim$ 40 mV. Thus the {\it gap-like} feature appears to close at $T_c$ in the manner of the superconducting gap in conventional superconductors in equilibrium.  However, such behavior is also expected from self-heating.  As $T_B$ increases, less voltage (heating power) is required to reach $T_c$ and for $T_B$$>$ $T_c$ the mesa is already in the normal state so no transition is possible. This discussion points out how misleading the heating effect can be.  Not only does the sharp upturn in $I(V)$ mimic the expected behavior of a tunnel junction at the gap voltage, but its disappearance above $T_c$ seems to (incorrectly) confirm its assignment as a superconducting energy gap.

\section{Summary and Conclusion}

In summary, we find that the occurrence of anomalously sharp conductance peaks in our data on Bi2212 mesas is an unexpected, counterintuitive consequence of heating.  Guided by the equilibrium spectra of single MCT junctions, these peaks are explained by a transition to the normal state through two-phase (normal and superconducting) coexistence, and it occurs at $V$$<$ $\mathit \Delta(T_B)/e$. The data also provide a measure of the severity of mesa heating by the figure-of-merit, $\beta$. Our uniform heating model, using values for $\mathit \Delta(T)$ and $\mathit \Gamma(T)$ from STS,~\cite{Pasupathy} shows quantitative consistency with the $N$-dependence of the conductance-peak {\it voltages} and the abrupt change in $\beta$ between $N$= 4 and $N$= 8. This, together with further analysis, gives strong evidence for us to conclude, unambiguously, that the sharp conductance peaks ($\beta$$<$ 0.15) are not a measure of the superconducting gap, but they are rather a consequence of two-phase coexistence due to strong self-heating.

Sharp conductance peaks are often seen in intermediate size mesas and they can be easily, but according to our analysis, incorrectly, assigned to  $\mathit \Delta(T)$, even though such  two-phase coexistence will mimic the closing of the {\it gap-like} feature at $T_c$. Thus the sharp peaks should not be used to make inferences about $\mathit \Delta(T)$ near $T_c$.  As a heating phenomenon, the sharp peak must disappear above $T_c$: this behavior is easily misinterpreted as a closing of $\mathit \Delta(T)$ and we reiterate the important consequences of such a false interpretation.  If a measurement seems to show that the superconducting gap closes at $T_c$ then it must necessarily lead to the conclusion that any pseudogap observed above $T_c$ is extrinsic to superconductivity. Such a conclusion would be at odds with other spectroscopic measurements.~\cite{Pasupathy, Campuzano, Eschrig}

The extreme difficulty of eliminating heating may imply the need to reinterpret some recent IJJ studies~\cite{Suzuki99, Krasnov05, Krasnov00, Katterwe, Bae08, Anagawa, Suzuki00} that generally exhibit small or non-existent DHF in $dI/dV$ and $\beta$ values of 0.03-0.15.~\cite{Suzuki99, Krasnov05, Krasnov00, Katterwe, Bae08} In view of the strong heating effects, one may also want to revisit the interpretation of the broad peaks seen in mesas near and above $T_c$. A comparison of the present work with other IJJ studies indicates that when  $\beta$$\geq$ 0.32 heating may not exceed $T_c$ in the mesa at $V$= 2$\mathit \Delta/e$.  Then the mesas are closer to equilibrium and the upturn in $I(V)$ is a measure of $\mathit \Delta$, albeit at an unknown $T_m$ somewhat higher than $T_B$.  In this case the voltage dependent $T_m$ ensures that $I(V)$ will never perfectly replicate any equilibrium $I(V)$ found by MCT or STS.  Another indicator, beyond $\beta$, of strong self-heating is the absence of a dip/hump feature, a characteristic signature of superconductivity in Bi2212.

While our conclusions reduce the usefulness of Bi2212 mesas for fundamental studies, it is important for the scientific community to recognize this limitation.

Work supported by UChicago Argonne, LLC, operator of Argonne National Laboratory, a U.S. Department of Energy Office of Science laboratory, operated under contract No. DE-AC02-06CH11357 and TUBITAK (Scientific and Technical Research Council of Turkey) project number 106T053.  L.O. acknowledges support from Turkish Academy of Sciences, in the framework of the Young Scientist Award Program (LO/TUBA-GEBIP/2002-1-17).

\end{document}